# STATUS OF THE EXPERIMENTS ON RADIATIVE BRANCH OF NEUTRON DECAY

Khafizov R.U.[a], Tolokonnikov S.V.[a], Solovei V.A.[b], Kolhidashvili M.R.[b]

[a] RRC Kurchatov Institute, 123182 Moscow, Russia
[b] Petersburg Nuclear Physics Institute, 188350 Gatchina, Russia

Abstract

This report is dedicated to the investigation of radiative neutron decay. The theoretical spectrum of radiative gamma quanta, calculated within the framework of the standard electroweak interaction model, is compared with our experimental value of branching ratio ( B.R. ) for radiative neutron decay. It is noted that the study of radiative branches of elementary particle decay occupies a central place in the fundamental problem of searching for deviations from the standard electroweak model. Particular attention is paid to analyzing the results of the experiment conducted at the FRMII reactor of the Technical University of Munich [1] in 2005.

**Introduction.**

Characteristics of the ordinary decay mode are currently measured with precision of tenths of a percentage point. Under these circumstances experimental data obtained by different groups of experimentalists can be reconciled only by taking into account the corrections calculated within the framework of the standard theory of electroweak interactions. This means that experimental research of the ordinary mode of neutron decay has exhausted its usefulness for testing the standard model. To test the theory of electroweak interaction independently it is necessary to move from the research of the ordinary decay branch to the next step, namely, to the experimental research of the radiative decay branch.

The radiative decay branch, where an additional particle, the radiative gamma quantum, is formed along with the regular decay products, has been discovered for practically all elementary particles. This has been facilitated by the fact that among the rare decay branches the radiative branch is the most intensive, as its value is proportional to the fine structure constant α and is only several percent of the intensity of the regular decay mode (in other words, the relative intensity B.R. of the radiative decay branch has the value of several hundredths of a unit.)

However, for the neutron this decay branch had not been discovered until recently. Our experiment conducted in 2005 at the FRMII reactor of Munich Technical University became the first experiment to observe this elementary process [1]. We initially identified the events of radiative neutron decay by the triple coincidence, when along with the two regular particles, beta electron and recoil proton, we registered an additional particle, the radiative gamma quantum

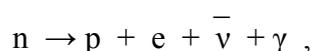

$$n \rightarrow p + e + \bar{\nu} + \gamma ,$$

and so could measure the relative intensity of the radiative branch of neutron decay B.R.= $(3.2 +- 1.6) \cdot 10^{-3}$ ( with C.L.=99.7% and gamma quanta energy over 35 keV; before this experiment we had measured only the upper limit on B.R. at ILL [2] ).

If we compare neutron decay with the well known radiative muon decay ( see ref. in [3] ),

the research of radiative neutron decay gives significant advantages. First, the presence of intensive cold neutron beams with intensity of $10^9$-$10^{10}$ n/s/cm$^2$ allows to research this rare mode of neutron decay in flight, without the stop target. In the muon case, the research of this rare decay mode requires the use of the stop target, the material of which creates a significant background of the bremsstrahlung gamma quanta. This background becomes significant at the low energies of the registered gamma-quanta, so at the present the muon B.R. is well researched only for the high energy part of the spectrum, with the energy of the registered gamma-quanta over several Mev. When researching the neutron decay on intensive beams, the stop target is absent, which allows to study the radiative gamma quanta of exceptionally low energies, up to the light photons. This, in turn, allows to experimentally test a phenomenon well known in theory - the infrared convergence. Secondly, neutron life time is several orders of magnitude longer than the longest-lived elementary particle of muon. This, along with the presence of the intensive cold neutron beams, allows to research the characteristics of the radiative neutron decay with much greater precision than it has been done before, even for the muon. Thus, conducting a precise experiment on the radiative neutron decay allows to hope to discover a deviation from the standard electroweak model.

The current status and perspectives for research of the neutron and muon radiative decay modes will be discussed in detail in a separate work. In this report we will only mention that while numerous experiments have been dedicated to the research of the radiative decay mode over the past few decades, the radiative decay mode of neutron in particular has been researched only recently and by only two experimental groups. Below we will analyze and compare the results of our experiment on radiative neutron decay to the attempt to identify radiative events in the decay of neutron by a NIST experimental group. We will also compare the results of our experiment with the experiment conducted by emiT group in NIST in 2000 to measure the ordinary decay mode [4].

In the first experiment we conducted at ILL to identify radiative events we can measure only upper limit on B.R. [2]. In the second experiment we conducted at FRMIII to measure the relative intensity of the radiative branch of decay, we received an average value for B.R. that exceeds the theoretical value we obtained within the framework of the electroweak model [5]. However, due to the presence of a significant experimental error, we cannot make any final conclusions about a deviation from the standard model. Thus, our next step is to measure the characteristics of the radiative neutron decay, and especially the B.R., with greater precision. Almost a year after we published our results on the internet [6] and at JETP Letters [1], another experimental group announced the discovery of the radiative neutron decay mode and cited its own value B.R. = $(3.13\pm0.34)\cdot10^{-3}$ with C.L.=68% and gamma quanta energy from 15 to 340 keV [7]. However, as we demonstrated later [8] the use of strong magnetic fields of several tesla prevents the identification not only of the radiative decay events, but also the events of ordinary neutron decay.

To summarize the main distinctions of our experiment [1] and compare them with the experiment conducted at NIST [7]. Both equipments are comparable in size and about 1 meter. The diagram of our equipment is given on Fig. 1. The size of our equipment is also comparable to the equipment used by emiT group at NIST in 2000 to measure coefficient D for the ordinary decay branch, an experiment to which we will also refer and with which we will compare our results [4].

Our spectrum of double coincidences of electron and recoil protons is given on Fig. 2 and its main characteristics are similar to an analogous spectrum published by emiT group ( see Fig.11 in ref. [4] ). This is an important point and requires a detailed comparison of our spectrum and the spectrum of the emiT group on double coincidences of electron and recoil proton in neutron decay. As Fig 2 and Fig 11 reference [4] show, there are two distinct peaks in the spectra. As we can see on Fig.2 and Fig. 11 in ref [4] two peaks are eliminated on both spectra. The first narrow peak is located in the middle of the time window where the electron opens the two equal backward and forward time windows.

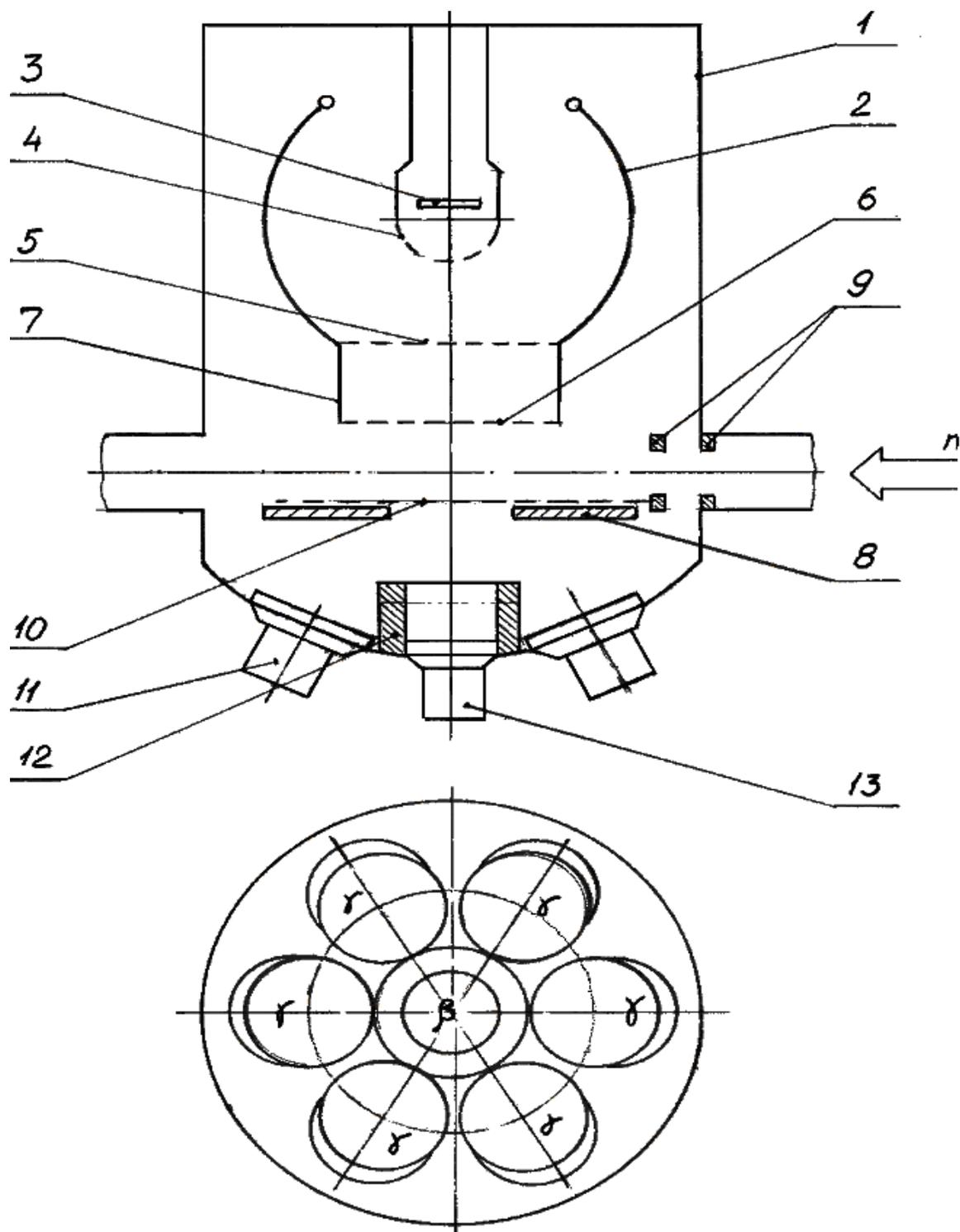

Fig.1. Schematic layout of the experimental set-up. (1) detector vacuum chamber, (2) spherical electrodes (at 18-20 kV) to focus the recoil protons on the proton detector (3) ( at ground potential ), (4) grid for proton detector (at ground potential), (5) & (6) grids for time of flight electrode, (7) time of flight electrode (at 18-20 kV), (8) plastic collimator (5 mm thick, diameter 70 mm) for beta-electrons, (9) LiF diaphragms, (10) grid to turn the recoil proton backward (at 22-26 kV), (11) six photomultiplier tubes for the CsI(Tl) gamma detectors, (12) lead cup, (13) photomultiplier tube for the plastic scintillator electron detector.

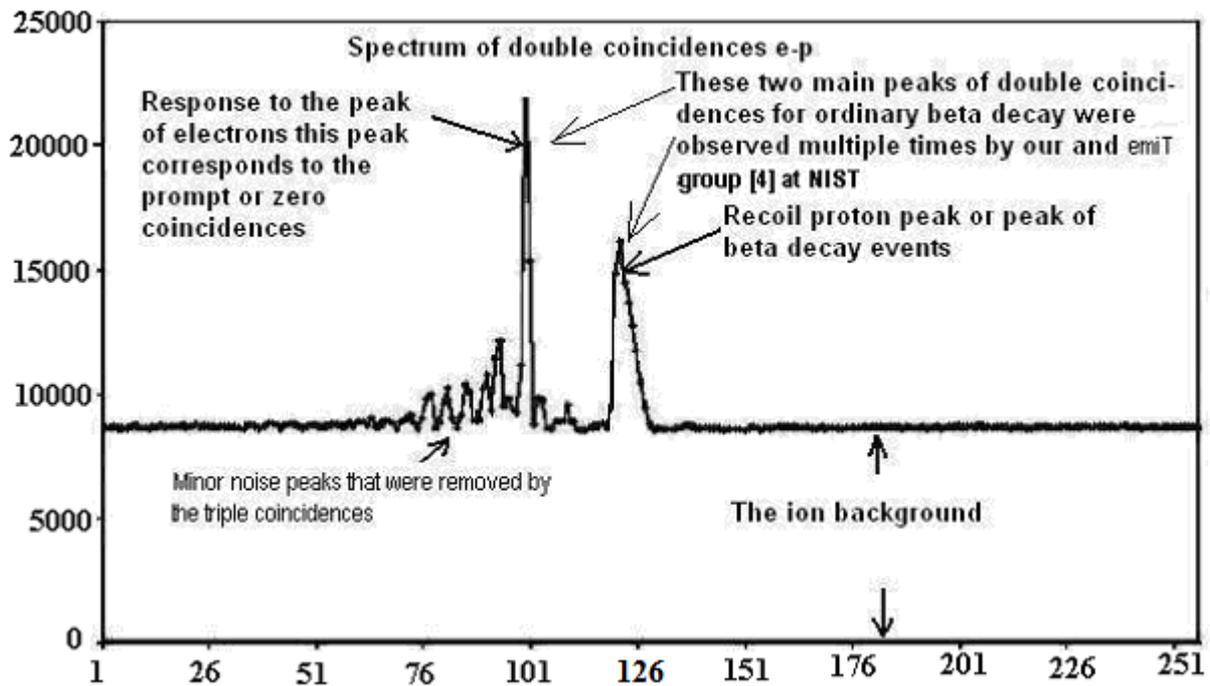

Fig.2. . Timing spectrum for e-p coincidences. Each channel corresponds to 25 ns. The peak at channel 99-100 corresponds to the prompt ( or zero ) coincidences. The coincidences between the decay electrons and delayed recoil protons (e-p coincidences) are contained in the large peak centered at channel 120 ( delay time is about 500 ns ). Minor noise peaks before the peak of zero coincidences were not stable during statistics collection, disappearing at nighttime and on weekends, when the noise in the electric circuits was minimal.

In essence, this first peak is the response to the registration of beta electrons in the electron detector channel. The second peak is located to the right of the first at about the same distance in both our spectrum and the spectrum emiT group from NIST. This peak is created by the decay protons which in turn were pulled out from the decay zone by the electric field. Thus, the events that got into this peak are beta decay events: we first register the beta electron in the electron channel and then after a slight delay register the recoil proton in the proton channel. It is namely the number of events in this peak that determines the total number of registered neutron beta-decays, necessary in our experiment to measure B.R.. In our equipment protons traveling from the decay zone cross a distance several times greater than in the emiT group equipment, but proton speed is also several times higher than in emiT group experiment. From this we deducted that proton delay time in relation to electrons in both experiments is comparable. The width of the proton peak in emiT group experiment is also comparable to the width of the proton peak in our spectrum of double coincidences, as this width is primarily determined by the size of the decay zone and proton speeds. In our experiment the thickness of decay zone was 3 cm while the diameter of the decay zone in the emiT experiment was slightly greater than 6 cm, which explains why the peak width in their case is roughly twice the size of ours. Here it is important to emphasize that initial speed of recoil protons is the same in both our the emiT group experiments and that the maximum speed is determined by the maximum kinetic energy equal to 750 eV. However, in our experiment protons leaving the decay zone immediately get picked up by the external electrostatic field and their final speed is determined by the difference of potentials between the focusing electrodes and the grounded grid of the proton detector (see Fig. 1 to compare it with Fig. 5 and 7 in ref. [4] ), which is why flight time to the proton detector was approximately equal in both our and the emiT group experiments, even though in our case the distance between the decay zone and the proton detector was several times greater than in the emiT group experiment.

Our double coincidences spectrum and the results of the emiT group are also similar in regard to the ion background. In both experiments the value of the background in the time window opened backward and located to the left of the sharp response peak (on Fig. 11 ref. [4] see value of background in window 1) is equal to the value of the ion background in the window opened forward (see Fig. 11 in ref. [4] see value of background in window 2) and located to the right of that peak. Most importantly, in both experiments the value of the background was comparable to the value of the beta-decay peak itself, which in turn means that in such experiments the background cannot be disregarded and that to determine the number of double coincidences it is necessary to distinguish this peak from the background.

To measure B.R. of the radiative neutron decay it is necessary to first identify the events of ordinary neutron decay and determine the number of double coincidences of beta electron with recoil proton $N_D$ and then to identify the events of radiative decay and determine the number of triple coincidences of electron, recoil proton and gamma-quantum $N_T$. That is exactly what we did in our experiment and obtained two spectra, the spectrum of triple coincidences, used to identify events of radiative neutron decay distinguishing the radiative peak from the background, and the double coincidences spectrum, used to identify events of ordinary neutron decay distinguishing beta decay peak from the background. Unfortunately the NIST team measuring B.R. for the radiative neutron decay did not publish the experimental spectrum of double coincidences [7], which meant that they also did not present the peak of ordinary neutron decay observed by both our team and the emiT group in the double coincidences spectra. The absence of the peak is obvious as the experiment on radiative decay conducted at NIST used strong magnetic fields with values of several tesla (while the emiT group also used a magnetic field, its value was negligible and was just several millitesla), and strong magnetic fields wash out the peak of ordinary decay so that it practically fades into the ion background and it becomes impossible to distinguish. As we noted above, both our group and the emiT group obtained a background value comparable to the value of the beta decay peak ( see Fig. 2 )**,** thus in principle it is possible to fit the experimental result to the B.R. value we calculated within the QED framework over 10 years ago by simply cutting out a time window in the ion background. As shown on Fig. 3, such manipulations with the ion background are absolutely unacceptable, as it is necessary to first observe the beta decay peak against the ion background on the double coincidences spectrum (see Fig. 2) and only then determine the number of ordinary beta decay events $N_D$ by subtracting the ion background from the obtained double coincidence spectrum [ 1 ].

Delays of recoil protons, published by the NIST group in [7] ( see Fig. 3 ) are greater than 10 microseconds and are in sharp contradiction to the elementary estimates (see Fig. 4 and 5) and to the experimental spectra, obtained by us and by the emiT group [4] at NIST. Thus, instead of identifying beta decay events, the NIST experiment on BR measurements identified events of double coincidences of the background ions with beta electrons, while as we pointed out earlier (see Fig. 3), experiment authors could use only the very long delay times, characteristic for ions, but not for protons. Here it is necessary to note that ions could be created inside the equipment (vacuum chamber, neutron guide) not only because of the high intensity or a neutron beam, but also because of radioactive emissions, created due to activation of media inside of the experimental set up as the gamma quanta energy of the emission substantially exceeds the energy of ionization.

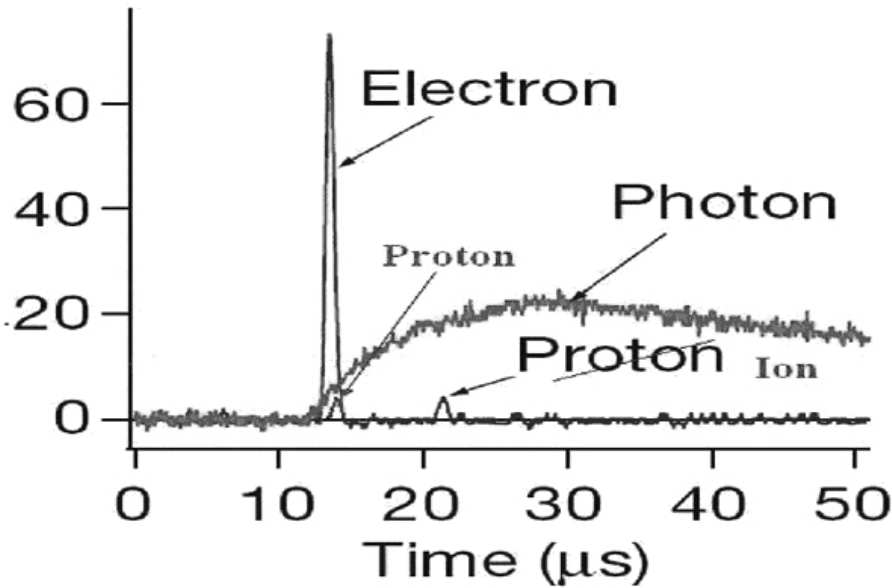

Fig.3. The photon line shows the shape of pulses from the gamma detector of the NIST experiment, built on avalanche diodes, and the electron line shows the shape of pulses from the combined electron-proton detector ( see ref. [7] ) The signal from the decay proton has to be delayed by less than one microsecond, which is why it is located at the base of the electron pulse and so cannot be registered by the combined electron-proton detector of NIST experiment. The pulses that are delayed by longer than 1 microsecond are pulses not from decay protons, as it was indicated in ref. [7], but rather from ions, formed in the decay zone ( see Fig. 2, which shows that the delay time between proton peak and the peak of prompt events is about 500 ns, and the ion background is significant because it has the same value as the proton peak ).

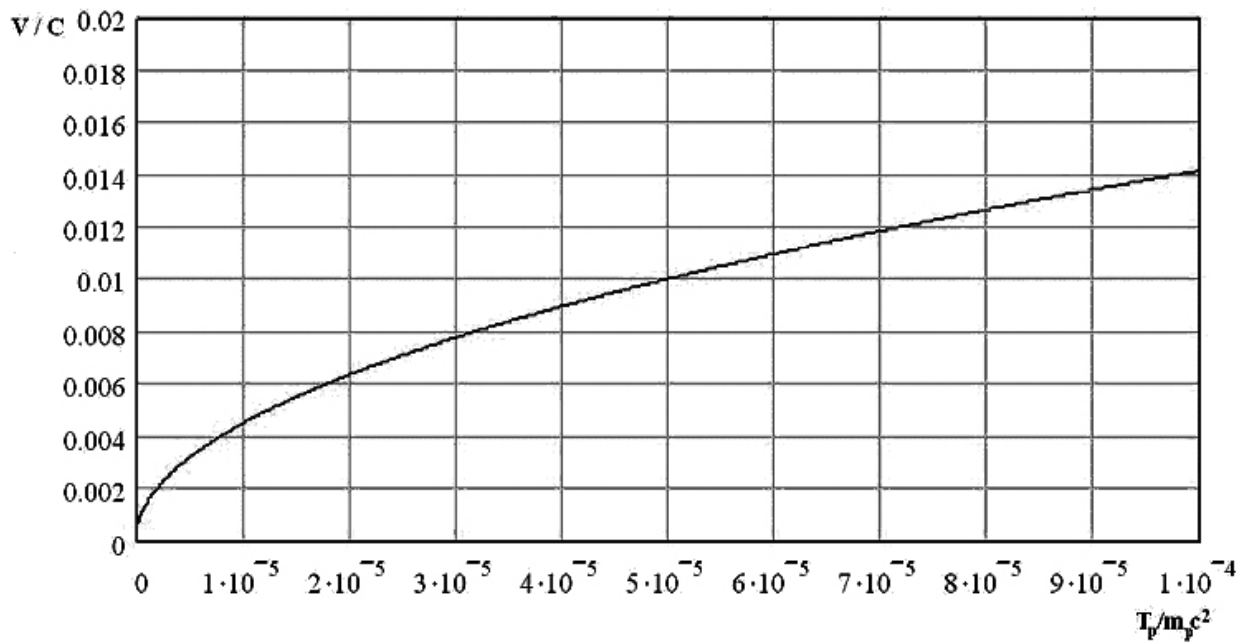

Fig.4. Relationship between the ratio of proton velocity v to speed of light c and the ratio of kinetic proton energy Tp to its mass $m_p c^2$

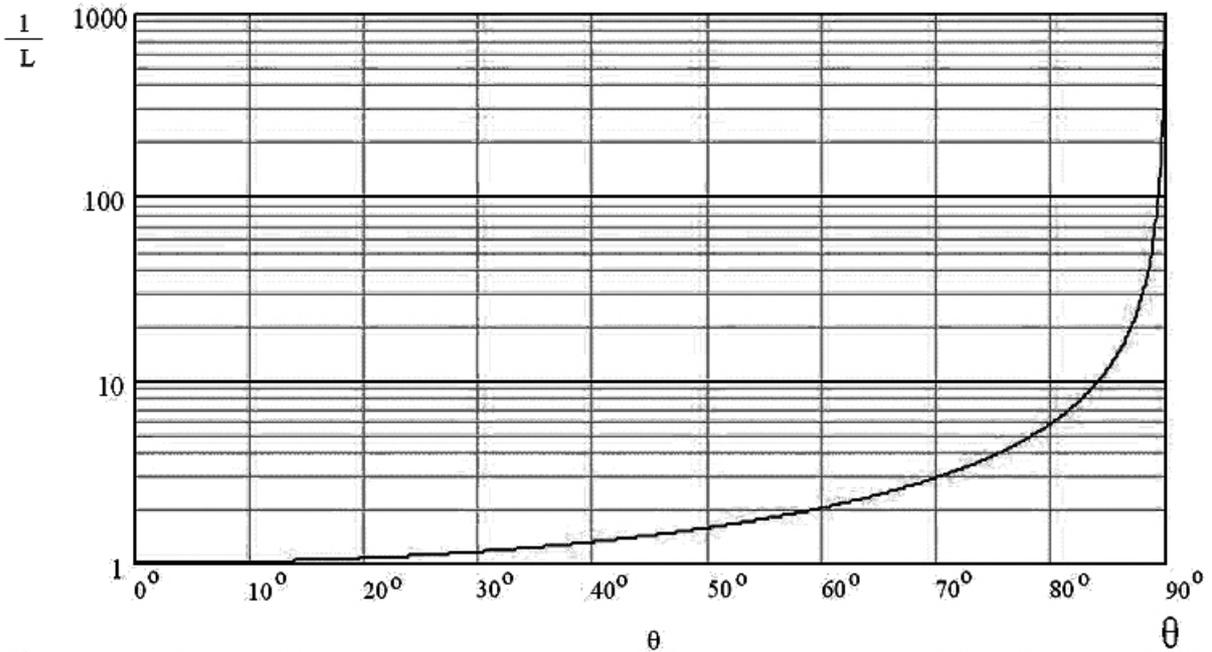

Fig.5. . Relationship between the ratio of the trajectory length $l$ of charged particles moving in the magnetic field to distance $L$ between the point of decay and the detector and the angle between the velocity of this particle and the direction of the magnetic field θ. In beta decay, the beta-electrons and the protons can fly out under any angle θ, therefore the magnetic field can increase the time of delay by several orders of magnitude only for a negligible portion of the charged particles. Even this negligible number of particles that flew out at an almost 90 degree angle to the direction of the magnetic field that coincides with the direction of the narrow neutron guide will most likely end up on the walls of the neutron guide rather than reach and hit the detector due to the presence of the strong electrostatic field.

Now we'll move to discuss the spectrum of triple coincidences, beta electron, recoil proton and radiative gamma quantum, created by radiative neutron decay. Our spectrum of triple coincidences is shown on Fig. 6, where the radiative peak is slightly shifted and located before the response to the beta-electron peak. The problem is that we observed the radiative peak not against a uniform horizontal background, but rather against a background of two peaks, which in their turn are responses to the two peaks in the spectrum of double coincidences in Fig. 2. Thus it is necessary to first distinguish the narrow radiative peak from this heterogeneous background using the method of response function. The number of events in this peak is what determines the number of radiative neutron decays, while the error arises from statistic fluctuations of the background, which can be seen on the horizontal parts of the spectrum of triple coincidences on Fig. 6. The final value of radiative events, namely the value of triple coincidences $N_T$, in our experiment with an error of 3σ was equal to $N_T$ = 360±180 events. After we obtained the number of beta-decay events forming the beta decay peak on the spectrum of double coincidences $N_D$ and calculated the geometric factor of the experimental equipment k [1], we obtained the B.R. = k · $N_T$ / $N_D$ value of B.R. = (3.2+-1.6) $10^{-3}$ ( with C.L.=99.7% and gamma quanta energy over 35 keV ). It is necessary to note that the average B.R. value is greater than the theoretical value, calculated within the framework of electroweak interaction. However, due to the significant error we cannot conclude that we observe a deviation from the standard model.

Coming back to the comparison of our results to the findings of the NIST experimentalists claiming to observe radiative decay events: as in the case with double coincidences, the NIST group has not published the initial experimental spectrum of triple coincidences. Instead, they demonstrate a single wide gamma peak, shifted by 1.25 microseconds before the beta-electrons. This value deviates from the estimates and the value of the leftward shift of the radiative peak from the response to the electron peak shown on our experimental spectra by several orders of magnitude. The location of the radiative peak is determined by the distance from the decay zone to detectors L and the speed of the registered articles. These distances in both experiments are roughly the same at about 1 meter and the presence of strong magnetic fields cannot lengthen spiral for the absolute majority of charged particles in this experiment even twofold (see Fig. 4), and the speed of beta electrons is comparable to the speed of light (see Fig. 5). Thus the radiative peak should be delayed by several tens of nanoseconds at most and definitely should not be delayed by microseconds.

A delay of 1.25 microseconds is huge in comparison to this estimate, and to obtain such a delay the size of the equipment should be hundreds of meters. Besides, during our experiment we did not obtain such a peak before the arrival of beta electron, and as can be seen from Fig. 6 the background before the peak responding to electron registration is horizontal and against it only negligible statistical fluctuations are visible. We can reconcile our spectra of triple coincidences with the one isolated peak observed at NIST [7] only if we assume that at NIST, the gamma-quanta were registered after the beta-electron ( see Fig. 6 ).

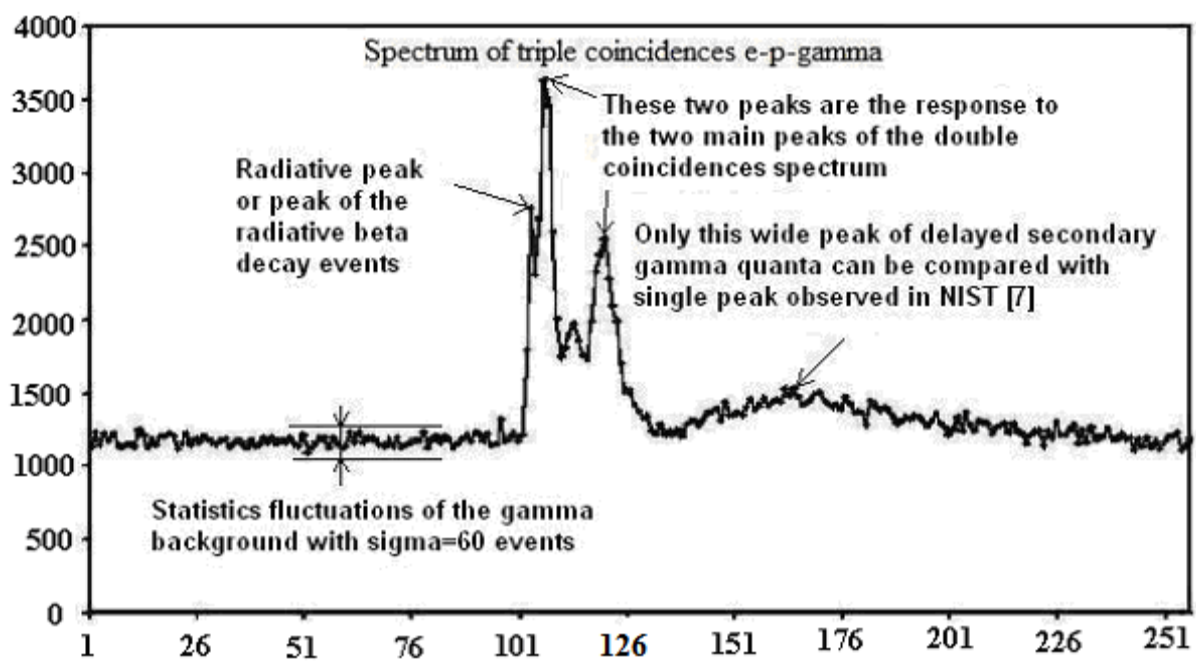

Fig.6. Timing spectrum for triple e-p-g coincidences. Each channel corresponds to 25 ns. In this spectrum, three main peaks in channels 103, 106 and 120 can be distinguished. The leftmost peak in 103 channel among these three main peaks is connected with the peak of radiative decay events. This peak was always stable, it never "migrated" to a different channel and it grew at a stable rate, regularly collecting the same number of events during the same stretch of time.

In this case the huge shift of 1.25 microseconds obtained by the NIST experimentalists fully corresponds to the shift of the wide peak we obtained in the spectrum of triple coincidences located to the right of the peak responding to beta electron registration, as this peak is formed by gamma quanta delayed in relation to the beta electron. In comparing these peaks it is also necessary to note that their widths are identical. A radiative peak, on the other hand, could not be so wide as to be comparable to 1 microsecond, as the width is primarily determined by the size of the observed decay zone and the speed of the gamma quanta, which are parts of a nanosecond. Thus our radiative peak is very narrow and occupies one channel, the minimal time interval accommodated by our electronic equipment. If we return to the double coincidences spectrum, we will see that the second wide beta decay peak is quite wide in both our experiment (see Fig. 2) and the emiT group experiment (see fig. 11 in reference [4]), and that its width in both experiments corresponds well to the elementary estimate, arrived at by simply dividing the size of the observed decay zone by the speed of the recoil proton. Even at recoil proton speeds substantially below the speed of light – the speed of radiative gamma quanta, the width of the proton peak of double coincidences [1, 4] is less than the width of the radiative peak obtained in the NIST experiment [7].

It is necessary to note that both our Western collaborators and the authors of the radiative decay experiment in NIST [9] perfectly understood that the only peak published in Nature, the wide peak, corresponds primarily to our wide peak in the triple coincidences spectrum formed by gamma quanta registered after the registration of beta electrons, as throughout our experiment we never observed a wide peak before the registration of beta electrons. Namely this understanding heightened the NIST experimentalists' interest in explaining the wide peak in our spectrum of triple coincidences. It is obvious both from the position and the width of the peak that it is not the radiative peak [10]. Instead, it is formed by delayed secondary gamma quanta emitted by the media inside of experimental chamber activated by the detected beta-electron. According to our estimates, the value of the peak corresponds namely to the emission intensity of the secondary radioactive gamma quanta with energies above 35 keV, created when the equipment environment is activated by beta electrons. The value of the peak in the NIST experiment was greater namely because they selected the lower energy limit for the registered gamma quanta, 15 keV.

**Conclusions.** From the arguments laid out above it follows that the authors of the NIST experiment did not identify the peak of beta decay in the spectrum of double coincidences. One reason for this is the use of strong magnetic fields in several tesla, as well as detectors with resolution several orders of magnitude worse than those used in our experiment. On the other hand, the main parameters of our spectrum of double electron-proton coincidences identifying the events of ordinary neutron decay fully coincide with an analogous spectrum published by emiT group in Phys. Rev. [4] ( see Fig. 2 and Fig. 11 in ref. [4]).

Unfortunately we cannot say same for another experiment measuring the radiative neutron decay conducted in NIST. Our results are in sharp contradiction to the results they published in Nature [7]. Particularly vexing is the authors' unsubstantiated assertion that they observe their only wide peak of gamma quanta before the registration of beta-electrons. Both the position and the width of this peak are located in sharp contradiction to both the elementary estimates [8], and the results of our experiment [1]. In the course of our entire experiment we did not observe such a wide gamma quanta peak in the triple coincidences spectrum, located before the arrival of electrons at a huge distance of 1.25 microseconds ( see Fig. 6 ). However, it is possible to reconcile our spectra of triple coincidences with the one isolated peak observed at NIST [7] if we assume that at NIST, the gamma-quanta were registered after the beta electrons. Only in this case does the NIST peak almost completely coincide with the peak we observed in the spectra of triple coincidences with the maximum in channel 163 ( see Fig. 6), both in terms of the huge delay of 1.25 microsecond and in terms of its huge width. It is obvious that this peak is in no way related to the radiative peak, as the gamma quanta are registered after the beta electrons. This peak is created by the delayed secondary radioactive gamma-quanta, arising from the activation of the media inside experimental chamber, which was the real object of the NIST experimentalists' observation. From this it follows that the authors of the NIST experiment knowingly mislead the physics community when they assert that the peak they observed is formed by the radiative gamma quanta registered before the registration of beta electron, when in reality their peak, shifted by 1.25 microseconds and based on comparable with this shift width is formed by delayed radioactive gamma quanta, the nature of which has nothing in common with the radiative gama quanta formed in radiative neutron decay, which is the subject of our research.

The main result of our experiment is the discovery of the radiative peak namely in the location and of the width that we expected. The location and the width of the radiative peak correspond to both estimates and the detailed Monte Carlo simulation of the experiment. Thus, we can identify the events of radiative neutron decay and measure its relative intensity, which was found to be equal B.R. = $(3.2+-1.6) 10^{-3}$ ( with C.L.=99.7% and gamma quanta energy over 35 keV ). At the same time, the average experimental B.R. value exceeds the theoretical value by 1.5 times. However, due to a significant error we cannot use this result to assert that we observe a deviation from the standard model. Therefore, our most immediate goal is to increase experiment precision, which we can improve by 10% according to estimates.

The authors would like to thank the President of RRC "Kurchatov Institute" Academician E.P. Velikhov, Prof. Yu. V. Gaponov and V. P Martemyanov for their support of our work. The research is conducted with the financial support of RFBR ( grant № 1a - 00517 ).